\documentclass[epj]{svjour}

\usepackage{graphicx}
\usepackage{amsmath}
\usepackage{amsfonts}
\usepackage{amssymb}

\begin{document}

\title{Castaing's instability in a trapped ultra-cold gas}
\author{J.N. Fuchs\thanks{email: fuchs@lkb.ens.fr},
O.\ Pr\'evot\'e\thanks{\emph{Present address:} Universit\'e de
Cergy-Pontoise,
D\'epartement de Physique, 5 mall Gay Lussac, Neuville sur Oise,
95031 Cergy Pontoise, France} \fnmsep \thanks{email:
olivier.prevote@voila.fr} \and  D.M. Gangardt
\thanks{email: gangardt@lkb.ens.fr}}

\institute{Laboratoire Kastler Brossel, D\'{e}partement de Physique de
l'ENS, 24 rue Lhomond, 75005 Paris, France}

\date{Received: date / Revised version: date}

\abstract{ We consider a trapped ultra-cold gas of (non-condensed)
bosons with two internal states (described by a
pseudo spin) and study the stability of a longitudinal pseudo 
spin polarization gradient. For this purpose, we
numerically solve a kinetic equation corresponding to a 
situation close to the experiment at JILA \cite{Cornell}.
It shows the presence of Castaing's instability of transverse 
spin polarization fluctuations at long wavelengths.
This phenomenon could be used to create spontaneous transverse spin waves.
\PACS{{03.75.Fi}{Phase coherent atomic ensembles; quantum condensation
phenomena}\and{51.10.+y}{Kinetic and transport theory of gases}\and{75.30.Ds}{Spin waves}}}

\maketitle

\section{Introduction}
Recent experiments \cite{Cornell} with a trapped ultra-cold (non condensed) Bose gas with
two internal states have shown the existence of
interesting relative population dynamics. Theoretical studies
\cite{Fuchs}\cite{OktelWilliams} making use of a pseudo spin
description for the internal states have explained this phenomenon
as pseudo spin oscillations due to the ``identical spin rotation
effect"(ISRE) \cite{LL1} which
appears when the temperature is low enough for the binary collisions to be in the
quantum regime. Alternatively, this mechanism can be understood as a 
``spin mean-field" \cite{Bashkin}. In a polarized system, this implies the existence of
low energy excitations of the
transverse spin polarization, named spin waves. The prediction of spin waves in dilute gases
\cite{Bashkin}\cite{LL2} as well as their observations in
H$\downarrow$ and helium \cite{spinwaves} goes back to
the 1980's. Shortly later, Castaing \cite{Castaing} showed
that a strong gradient of longitudinal spin polarization is unstable
with respect to transverse fluctuations. His study focused on
homogeneous polarized $^3$He gas 
and assumed that the spin waves were in the hydrodynamic regime.
The purpose of this article is to provide a quantitative study of
the existence of Castaing's instability in an inhomogeneous system
relevant to the experimental situation at
JILA \cite{Cornell}. In this experiment, neither the
hydrodynamic nor the collisionless regime for the spin oscillations
are reached. This work is motivated by the recent contribution of Kuklov and Meyerovich
\cite{Meyerovich}, who were the first to suggest and study the existence of
Castaing's instability in this context. For completeness, one should
be aware of the numerical studies of Castaing's instability in 
polarized Fermi fluids, which are based on the solution of the
Leggett equations rather than the full kinetic equation \cite{Ragan}. 

\section{Kinetic equation; Leggett equations}
The physical situation we consider is close to that of the experiment at JILA
\cite{Cornell}: $^{87}$Rb atoms (bosons) with
two hyperfine states of interest (denoted by $1$ and $2$) are confined
in an axially symmetric magnetic trap elongated
in the $Ox$ direction. The temperature $T$ is about twice the critical
temperature for Bose-Einstein condensation, so
that the gas is non-degenerate (Boltzmann gas). However, the de
Broglie thermal wavelength is much larger than the scattering length
so that the collisions occur in the quantum regime. It
is convenient to consider the pseudo spin associated
with the two hyperfine states $1$ and $2$ (the basis in spin space is denoted
by $\{ e_{\bot,1}$; $e_{\bot,2}$; $e_{\parallel}\}$). Initially the spin
polarization is longitudinal (i.e. along the ``effective external
magnetic field") and has a strong spatial gradient. For example,
on the left (resp. right) of the trap center, the cloud of
atoms is mostly in state $1$ (resp. $2$). This might be achieved, for example, by
preparing a cloud of atoms in state $1$ and a cloud of atoms in state
$2$ separated by a sharp optical potential barrier at the center of
the magnetic trap. After removal of the optical barrier,
as the two clouds mix, a strong longitudinal spin polarization
gradient appears in the region of overlap. 
If it is strong enough, one should be able to observe the
appearance of a large transverse component of the spin polarization as
a result of Castaing's instability.

To study this system, we write an effective one dimensional kinetic
equation \cite{Fuchs} in terms of a local density in phase space
$f(x,p,t)$ and (pseudo) spin density $\mathbf{M}(x,p,t)$, where $x$ is the
position, $p$ the momentum, and $t$ the time. The one dimensional
description for an elongated system is justified by the time scale
separation associated with radial and axial motions characterized
by frequencies $\omega_{rad}$ and $\omega_{ax}$ respectively. The time
scales differ by more than an order of magnitude leading to an effective dynamical
averaging over radial dynamics as described in Ref. \cite{Fuchs}\cite{OktelWilliams}.
We define the local density $n$ and spin polarization $\mathbf{m}$ by
\begin{eqnarray}
n(x,t)&=&\int dp \, f(x,p,t) \nonumber \\
\mathbf{m}(x,t)&=&\int dp \, \mathbf{M}(x,p,t)
\label{density}
\end{eqnarray}
The local density of atoms in state $1,2$ is then given by $n_{1,2}=(n \mp
m_{\parallel})/2$, defining $m_{\parallel}$ as the longitudinal component 
of the spin polarization.
We also denote $g_{ij}=4\pi\hbar^2a_{ij}/m$ as the coupling constants associated to
the different scattering lengths $a_{ij}$ where $i$ and $j=1,2$
($a_{21}=a_{12}$ and $m$ is the mass of the particles).

We now review the different terms entering the kinetic
equation (for details, see Ref. \cite{Fuchs}).
The force acting similarly on both internal states contains three terms: the
magnetic trap, the mean-field, and the Stern-Gerlach
force (associated to a gradient of ``effective external magnetic
field'' $\boldsymbol{\Omega}$, see below). The magnetic trap force is
dominant, so the associated potential energy is given by
\begin{equation}
\frac{V_1(x)+V_2(x)}{2}=\frac{1}{2} m \omega_{ax}^2 x^2
\end{equation}
where $V_{1}$ and $V_{2}$ are the magnetic trapping potentials acting on states $1$ and $2$.
The differential force can be described in the pseudo spin picture by
an ``effective external magnetic field'' $\boldsymbol{\Omega}$ which contains
two contributions: one due to the differential Zeeman effect and one
due to the differential mean field, so that
\begin{equation}
\hbar \boldsymbol{\Omega}(x)=[V_2(x)-V_1(x)+(g_{22}-g_{11})n(x)/2]\mathbf{e}_{\parallel}
\label{eemf}
\end{equation}
where $\mathbf{e}_{\parallel}$ is a unit vector in
the longitudinal direction.
Following Ref. \cite{Cornell}, we have assumed that $2g_{12} \simeq
g_{11}+g_{22}$ for simplicity. The average value over the sample of the ``effective external
magnetic field'' is removed by going to a uniformly rotating frame (Larmor frame).
It is also crucial to consider the ISRE which manifests itself as a ``molecular field'' or ``spin
mean field'' $g_{12}\, \mathbf{m}(x,t)/2$, 
which adds to the ``effective external magnetic field'' (\ref{eemf}).
Finally, the collision integral is treated in a relaxation-time
approximation with a time $\tau$ of the order of the average time between
collisions. The kinetic equation is written as
\begin{eqnarray}
\partial_{t}f+\frac{p}{m}\, \partial_{x}f-m\omega_{ax}^2x\, \partial_{p}f
&\simeq&-(f-f^{eq})/\tau \label{f} \\
\partial_{t}\mathbf{M}+\frac{p}{m}\, \partial_{x}\mathbf{M}-m\omega_{ax}^2x\, \partial_{p}\mathbf{M}%
&-&(\boldsymbol{\Omega}+\frac{g_{12}\mathbf{m}}{2\hbar})\times\mathbf{M}\nonumber \\
&\simeq&-(\mathbf{M}-\mathbf{M}^{eq})/\tau \label{kinetic}
\end{eqnarray}
where $f^{eq}$ and $\mathbf{M}^{eq}$ are local equilibrium phase space densities.

In the hydrodynamic regime close to local equilibrium, the above equations reduce to
the counterpart of the Leggett equations \cite{Leggett} in the case of a
Boltzmann gas \cite{LL2}, namely
\begin{eqnarray}
\partial_{t}\mathbf{m}+\partial_{x}\mathbf{j}&=&\boldsymbol{\Omega}\times\mathbf{m} \nonumber\\
\partial_{t}\mathbf{j}-(\boldsymbol{\Omega}+\frac{g_{12}\mathbf{m}}{2\hbar})\times\mathbf{j}&+&\frac{k_BT}{m}
\partial_{x}\mathbf{m}+\omega_{ax}^2x\mathbf{m}\nonumber\\
&\simeq&-\frac{\mathbf{j}}{\tau} \label{Leggett}
\end{eqnarray}
where $\mathbf{j}(x,t)$ is the spin polarization 
current along $Ox$. These equations are also valid
in the collisionless regime close to global equilibrium \cite{Leggett}.
We define $\mu=g_{12}n(0)\tau/2\hbar$ as the dimensionless parameter
which characterizes the strength of the ISRE and 
denote $D=k_{B}T\tau/m$ as the spin diffusion coefficient.

\section{Castaing's instability}
Studying polarized $^3$He gas, Castaing \cite{Castaing} noticed
that a sufficiently strong longitudinal spin polarization
gradient is unstable against transverse long wavelength
fluctuations. This result was obtained by analyzing a uniform gas
($\omega_{ax}=0$) with a uniform precession frequency (so
that $\boldsymbol{\Omega}=0$ in the Larmor frame) and
by assuming a time independent spin gradient.
Using the Leggett equations (\ref{Leggett}), Castaing studied the stability
of this system with respect to the transverse fluctuations. He assumed 
a small plane wave perturbation around a slightly non-uniform stationary
solution such that
\begin{eqnarray}
\mathbf{m}(x,t)&=&m_{\parallel}^0(x)\mathbf{e}_{\parallel}+\delta\mathbf{m}e^{i(kx-\omega t)} \nonumber \\
\mathbf{j}(x,t)&=&-D\partial_{x}m_{\parallel}^0(x)\mathbf{e}_{\parallel}+\delta\mathbf{j}e^{i(kx-\omega t)}
\end{eqnarray}
For circularly polarized transverse spin waves 
in the presence of a strong gradient of longitudinal spin polarization, 
one obtains the following spectrum for the hydrodynamic regime ($\partial_{t}\mathbf{j}\simeq 0$)
\begin{equation}
\omega=\frac{D}{1+(\mu m_{\parallel}^0/n)^{2}} \left( \mu m_{\parallel}^0/n-i \right) \left( k^2-\mu k
(\partial_{x}m_{\parallel}^0)/n \right)
\label{dispersion}
\end{equation}
Here, a mode with wave vector $k$ is unstable if the mode frequency $\omega$ has a
positive imaginary part $\omega_{I}$, which leads to instability when
\begin{equation}
k^2 < \mu k (\partial_{x}m_{\parallel}^0)/n
\label{condition}
\end{equation}

The goal of this paper is to study Castaing's instability 
in a trapped ultra-cold gas. Using a numerical simulation,
described in the next section, we show that an instability does
occur. The situation we study is richer than the one considered in
Castaing's original work in several aspects: 
(i) with the full kinetic equation (instead of Leggett equations) we can describe spin
oscillations of large amplitude, (ii) we are not limited to the
hydrodynamic regime (for example, regimes between hydrodynamic 
and collisionless are included), (iii) the
longitudinal spin polarization gradient is not assumed to be constant, 
and (iv) we can also include a possible
gradient of the precession frequency (i.e. the 
``effective external magnetic field'' is position dependent).

\section{Numerical simulation}
The kinetic equation (\ref{f},\ref{kinetic}) is solved numerically by
propagating in time the initial distribution in a discretized
phase-space using the Lax-Wendroff method (see e.g. \cite{numrec}).
\begin{figure}[ptb]
\begin{center}
\includegraphics[ height=2.5in] {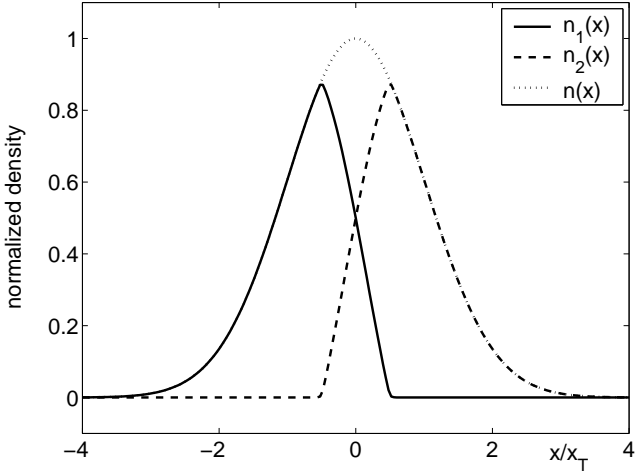}
\end{center}
\caption{Initial density of atoms in state $1$ (solid line) and $2$
(dashed line) giving a strong gradient of longitudinal spin polarization
near the center of the trap ($\partial_x
m_{\parallel}^0/n(0)=2/x_T$). The total density is
also plotted (dotted line).}
\label{initial_n1_n2}
\end{figure}
We assume that the density in phase space is initially at equilibrium
$f=f^{eq}\propto \exp(-x^2/2x_{T}^2-p^2/2p_{T}^2)$, with $x_{T}=\sqrt{k_{B}T/m\omega_{ax}^2}$ and
$p_{T}=\sqrt{mk_{B}T}$. As the equilibrium distribution does not
evolve in time under the action of (\ref{f}), the local density is
constant and we concentrate on the evolution of $\mathbf{M}$ described
by the kinetic equation (\ref{kinetic}).
The initial spin density distribution $\mathbf{M}$ is taken as the
product of the Maxwell-Boltzmann equilibrium distribution along the longitudinal axis
and a function which equals $-1$ to the left of the trap center
($x<-x_T$), $+1$ to
the right ($x>x_T$) and has a constant positive slope through the center
($-x_T<x<x_T$). The resulting initial density of atoms in state $1$
and $2$ is shown in Figure \ref{initial_n1_n2}. We introduce a  
small initial transverse perturbation in order to start the
instability. Without loss of
generality, the perturbation can be written as 
a plane wave (with spatial frequency of order $1/x_T$) multiplied by the
Maxwell-Boltzmann equilibrium distribution and divided by a number
$\mathcal{N} \gg 1$ (typically between $10^3$ and $10^6$):
\begin{equation}
M_{\bot,1}(x,p)=f^{eq}(x,p)\cos(\pi x/x_{T})/\mathcal{N}
\end{equation}

Parameters used in the simulation are taken from Ref. \cite{Cornell}.
The axial trapping frequency is $\omega_{ax}/2\pi=7$~Hz and the time
between collisions $\tau \sim 10$~ms, so that $\omega_{ax}\tau \sim 0.5$. The
``effective external magnetic field'' is taken to be an inverted
Gaussian of depth $\delta \Omega$ ($|\delta \Omega|/\omega_{ax}$ is
varied between $0$ and $2$) and half-width $x_T$. The density at the center of the trap is
$n(0)=1.8\times 10^{19}$~m$^{-3}$. The initial spin polarization
gradient near the center of the trap is varied 
and typically $|\partial_x m_{\parallel}|/n(0) \sim 2/x_{T}$. 
For $^{87}$Rb atoms, with $a_{12}=5.2 \times {10}^{-9}$~m and
$T=0.6$~$\mu$K, the
ISRE parameter is $\mu \sim 5$ . With these
parameters, an instability can already be observed in our 
simulations but is probably too weak to be
detected experimentally. To enhance the effect, we take a scattering
length $5$ times smaller
($a_{12}\rightarrow a_{12}/5$) and an axial
trap frequency $20$ times smaller ($\omega_{ax}\rightarrow
\omega_{ax}/20$), keeping a constant density at the center of the
trap. We discuss these matters further in the next section.

As the phase space is modeled by a discrete grid with finite size,
it is important to distinguish between a physical
instability and a numerical one. This can be easily done as only the
latter will depend on the grid spacing. 
One can always get rid of a numerical
instability by choosing a sufficiently tight grid. 

\begin{figure}[ptb]
\begin{center}
\includegraphics[ height=2.5in] {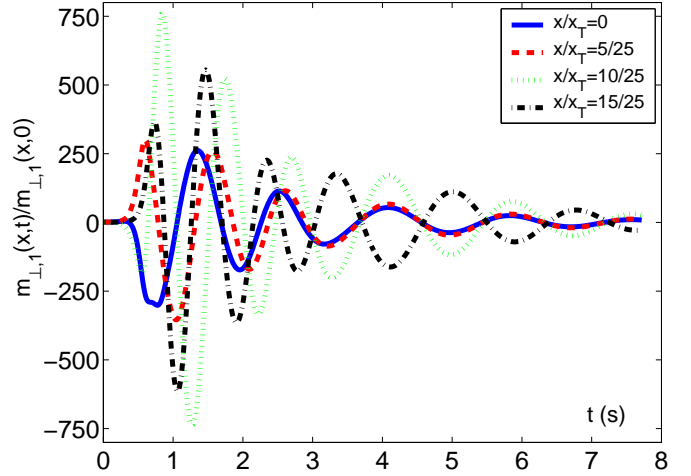}
\end{center}
\caption{Time evolution of the (normalized) transverse spin
polarization $m_{\bot,1}(x,t)/m_{\bot,1}(x,0)$ at different positions
in the trap: $x/x_T=0$ (solid line); $5/25$ (dashed line); $10/25$ (dotted
line); $15/25$ (dashed-dotted line). For this simulation $\mu\simeq22$,
$\omega_{ax}\tau\simeq 0.6$, $\delta \Omega/\omega_{ax}=2$,
$\partial_x m_{\parallel}^0/n(0)=2/x_T$ and $\mathcal{N}=10^3$.}
\label{time_evol}
\end{figure}
\begin{figure}[ptb]
\begin{center}
\includegraphics[ height=2.5in] {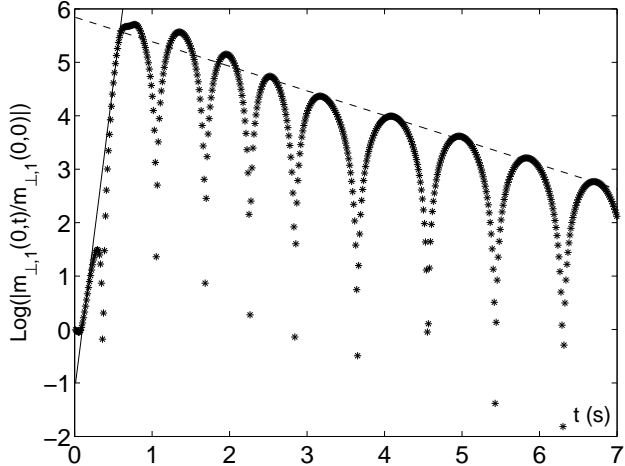}
\end{center}
\caption{Time evolution of the logarithm of the (normalized) transverse spin
polarization $m_{\bot,1}(0,t)/m_{\bot,1}(0,0)$ at the center of the
trap. The two lines are plotted to visualize the
exponential rise of the instability (full line) and then the
exponential decay of the spin wave (dashed line). The parameters have
the same values than for the simulation of Fig. \ref{time_evol}.}
\label{log}
\end{figure}

Results of the simulation are shown in Figures \ref{time_evol} and \ref{log}.
The time evolution of the transverse spin polarization is
plotted in Figure \ref{time_evol} for different positions near
the center of the trap, where the longitudinal spin polarization gradient
is most pronounced. The instability is clearly visible, owing to a
large enhancement of the transverse spin polarization 
by a large factor comparable to $\mathcal{N}$.
Figure \ref{log} shows the logarithm of the time evolution of the
(normalized) transverse spin polarization at the center of the trap.
In this representation, one can clearly distinguish the
exponential rise of the envelope (after a short delay of order of the time
between collisions) and its subsequent exponential
decay.

\section{Discussion}
The initial rise of the transverse spin polarization shows that an
instability indeed occurs. As already stated, the results presented here were
obtained for a scattering length $a_{12}$ and trap
frequency $\omega_{ax}$ that were smaller than in the current experiment
at JILA \cite{Cornell} (by factors of $5$ and $20$ respectively). 
This was done to enhance the effect of
the ``spin mean-field" (to favor the instability) by increasing the
ISRE parameter $\mu$\footnote{This could be done experimentally by using a Feshbach
resonance to decrease the scattering
length, as $\mu \sim \lambda_{T}/a_{12}$ where $\lambda_{T}$ is the de
Broglie thermal wavelength.}. 
As will be seen below, we have to keep $\omega_{ax} \tau \lesssim 1$,
which implies scaling the axial trap frequency with the square of the scattering 
length since $\tau^{-1} \propto a_{12}^2$. This procedure for
enhancing the ISRE insures that the gas remains non-degenerate.

The phenomenon observed in Fig. \ref{time_evol} 
and \ref{log} may be broken down into the following four steps: 

I.) During a time of the order of the time between
collisions $\tau\sim 250$~ms a hydrodynamic description is not valid (the spin
polarization current, which is a fast variable, has not reached its
stationary value yet). Figures \ref{time_evol} and
\ref{log} show that the transverse spin polarization does not
evolve significantly.

II.) Then, as shown by the solid line
in Figure \ref{log}, the envelope of the transverse spin polarization
rises exponentially (the coefficient in the exponential
$\omega_{I}$ is almost constant in time). This is a characteristic of an
instability. The value of $\omega_{I}$ (as compared to the formula
obtained by Castaing, see equation (\ref{dispersion})) is modified by
the presence of the trap 
. An estimate of the time
needed for the instability to develop (see Ref. \cite{Meyerovich}) is
given by $t_{inst}\sim L_{m}^2/D\simeq 1/\omega_{ax}^2\tau\sim 1$~s
(where $L_{m}$ is the characteristic size of the longitudinal spin polarization
gradient;  $L_{m}\simeq x_{T}$ in a typical simulation). It reproduces
correctly the order of magnitude of the observed maximum of transverse spin
polarization in the simulation. The instability can develop only if
the spin polarization current has reached its stationary value, so
that $\tau < t_{inst}$ which implies $\omega_{ax}\tau \lesssim 1$.

III.) The imaginary part of the frequency $\omega_{I}$ varies slowly
in time because the gradient of longitudinal spin polarization is not
constant but decays as a result of both spin diffusion and the
presence of the trap. The change of sign of $\omega_{I}$ marks the end
of the exponential grow. Once $\omega_{I}$ is negative, the transverse
spin polarization decays as an ordinary damped spin wave.

IV.) The longitudinal spin polarization gradient finally decays on
a time scale of the order of the diffusion time $t_{diff}\sim L^2/D \sim
4$~s  where $L$, is the size of the cloud ($L/x_{T} \sim 2$). As
emphasized by Kuklov and Meyerovich \cite{Meyerovich}, if
$L>L_{m}$, the instability develops faster than the longitudinal spin
polarization relaxes, in accordance with the results of our simulation. Once
the longitudinal spin polarization is zero, $\omega_{I}$ ceases to
evolve in time, and consequently the envelope of
the transverse spin wave decreases exponentially (dashed line in
Figure \ref{log}). This happens for $t\gtrsim t_{diff}$.
The frequency of the spin wave is not completely determined by the real part
of the mode frequency in equation (\ref{dispersion}), as the gradient of
external precession frequency, $\partial \boldsymbol{\Omega}$, 
and the presence of the trap are not taken
into account. Actually, its order of magnitude (at the center of the trap) is
given by the external precession frequency
$\delta \Omega = 2\omega_{ax} \simeq 2\pi \times
0.7$~Hz. We have checked that when $\delta\Omega/\omega_{ax}=0$, 
the transverse spin polarization at the
center of the trap does not oscillate.

The criterion for an instability (see equation
(\ref{condition})) is qualitatively verified at the
center of the trap if one takes into account the fact
that wave vectors $k$ are limited by the presence of the
trap. This forces $k>2\pi/L$ and the criterion becomes
\begin{equation}
\mu \frac{|\partial_x m_{\parallel}^0|}{n(0)} >\frac{2\pi}{L} \sim \frac{\pi}{x_T}
\end{equation}
Using an initial longitudinal spin polarization gradient $\partial_x
m_{\parallel}^0/n(0)=2/x_T$ and $L\sim 2x_T$ implies that the ISRE parameter $\mu$
should be larger than $\sim 2$. In the simulation, we found an
instability threshold at $\mu \simeq 4$. Obtaining a
significant instability requires a much larger value of the ISRE parameter.

We now discuss the relevance of this criterion for 
one of the experiments done  at JILA (see the first article 
of Ref. \cite{Cornell}), where spatial separation of the
two internal states was observed as a result of an initial 
$\pi/2$ rf pulse. The maximum longitudinal spin
polarization occured at the maximum of spin state separation 
and can be estimated as $|\partial_xm_{\parallel}|/n(0)\sim
1/x_T$ (at $x \sim \pm x_{T}/2$). Since the value of 
the ISRE parameter is $\mu \sim 5$, our numerically obtained 
criterion ($\mu |\partial_x m_{\parallel}^0|/n(0)
\gtrsim 8/x_T$) 
predicts no instability. Whether it is experimentally 
feasible in practice to reach transient total (or nearly total) separation
of the two spin states, so that, when re-mixing, the longitudinal 
spin polarization gradient would be strong enough
for Castaing's instability to develop, is not clear.
A very strong initial gradient of longitudinal spin
polarization with $\mu \sim 5$ may not be enough to start the instability because the
spin diffusion is very efficient in decreasing a strong gradient. One would
have to maintain a strong longitudinal spin polarization gradient in
order to create an instability with the current experimental value of $\mu$.

\section{Conclusion}
Our numerical simulation confirms the possibility of observing Castaing's instability in
a trapped ultra-cold gas with two internal states, as proposed by
Kuklov and \hyphenation{Me-ye-ro-vich}Meyerovich
\cite{Meyerovich}. We solved a one dimensional kinetic equation
numerically, and were therefore able to include effects that 
are beyond the usual treatment of Castaing's
instability in terms of a small amplitude hydrodynamic description.
We argue that Castaing's instability was
probably not relevant for the previous experiments done at JILA
\cite{Cornell} as both the ISRE parameter $\mu$ and the
longitudinal spin polarization gradient were too small. This does not
preclude observation of the instability in future experiments if
relevant parameters like the trapping frequency and the scattering
length are chosen appropriately.
We suggest the use of Castaing's instability as a way of creating
spontaneous transverse spin waves as a result of a strong initial 
longitudinal spin polarization gradient. Our calculations are also
valid for a non-degenerate gas of fermions (see Ref. \cite{Fuchs}), 
where similar effects could be observed, 
in a case where $g_{11}=g_{22}=0$ 
and the ``spin mean-field'' changes sign.

\bigskip
\begin{acknowledgement}
We are grateful to Franck Lalo\"{e} for many useful discussions and to
Chris Bidinosti for reading the manuscript.
Le Laboratoire Kastler Brossel (LKB) est une Unit\'{e} Mixte de
Recherche du CNRS (UMR 8552), de l'ENS et de l'Universit\'{e} Pierre et
Marie Curie (Paris).
\end{acknowledgement}


\bigskip

\begin{thebibliography}{}
\bibitem {Cornell}H.J.\ Lewandowski, D.M.\ Harber, D.L.\ Whitaker and
E.A.\ Cornell, Phys. Rev. Lett. \textbf{88}, (2002) 07403; J.M. McGuirk, H.J.\ Lewandowski, D.M.\ Harber, T.
Nikuni, J.E. Williams and E.A.\ Cornell, Phys. Rev. Lett. \textbf{89}, (2002) 090402.

\bibitem{Fuchs}J.N. Fuchs, D. Gangardt and F. Lalo\"{e},
Phys. Rev. Lett. \textbf{88}, (2002) 230404.

\bibitem{OktelWilliams}M.\"{O}. Oktel and L. Levitov,
Phys. Rev. Lett. \textbf{88}, (2002) 230403;
J. Williams, T. Nikuni and C.W. Clark, Phys. Rev.
Lett. \textbf{88}, (2002) 230405.

\bibitem {LL1}C.\ Lhuillier and F.\ Lalo\"{e}, J.\ Physique \textbf{43}, (1982) 197.

\bibitem {Bashkin}E.P.\ Bashkin, JETP Lett. \textbf{33}, (1981) 8.

\bibitem {LL2}C.\ Lhuillier and F.\ Lalo\"{e}, J.\ Physique \textbf{43},
(1982) 255.

\bibitem {spinwaves}B.R.\ Johnson, J.S.\ Denker, N.\ Bigelow, L.P.\ L\'{e}vy,
J.H.\ Freed, and D.M.\ Lee, Phys.\ Rev.\ Lett.\ \textbf{52}, (1984) 1508; W.J.\ Gully and W.J.\ Mullin, Phys.\
Rev.\ Lett.\ \textbf{52}, (1984) 1810; G. Tastevin, P.J. Nacher, M. Leduc, F. Lalo\"{e}, J. Phys. Lett.
\textbf{46}, (1985) 249.

\bibitem {Castaing}B.\ Castaing, Physica \textbf{126B}, (1984) 212.

\bibitem {Meyerovich}A.\ Kuklov and A.E.\ Meyerovich, 
Phys. Rev. A \textbf{66}, (2002) 023607.

\bibitem {Ragan} R.J. Ragan and D.M. Schwarz, J. Low
Temp. Phys. \textbf{109}, (1997) 775; R.J. Ragan and R.W. Weber, J. Low
Temp. Phys. \textbf{118}, (2000) 167. 

\bibitem {Leggett}A.J.\ Leggett, J.\ Phys.\ C \textbf{3}, (1970) 447.

\bibitem {numrec}W.H. Press, B.P. Flannery, S.A. Teukolsky and W.T.
Vetterling, \emph{Numerical Recipes} (C.U.P., 1986), chapter 17.

\end{thebibliography}
\end{document}